\documentclass[prl,twocolumn,floatfix,superscriptaddress]{revtex4-1}
\usepackage{dcolumn,amsmath}
\usepackage{graphicx}
\usepackage{bm}
\usepackage{hyperref}
\usepackage{xcolor}

\usepackage[utf8]{inputenc}
\usepackage[T1]{fontenc}

\usepackage{tabularx}
\usepackage{braket}

\begin{document}

\title{Effect of the neutron quadrupole distribution in the TaO$^+$ cation}

\author{Gleb Penyazkov}
\email{glebpenyazkov@gmail.com}
\affiliation{Petersburg Nuclear Physics Institute named by B.P.\ Konstantinov of National Research Center ``Kurchatov Institute'' (NRC ``Kurchatov Institute'' - PNPI), 1 Orlova roscha, Gatchina, 188300 Leningrad region, Russia}
\affiliation{Saint Petersburg State University, 7/9 Universitetskaya nab., St. Petersburg, 199034 Russia}

\author{Leonid V.\ Skripnikov}
\email{skripnikov\_lv@pnpi.nrcki.ru,\\ leonidos239@gmail.com}
\homepage{http://www.qchem.pnpi.spb.ru}
\affiliation{Petersburg Nuclear Physics Institute named by B.P.\ Konstantinov of National Research Center ``Kurchatov Institute'' (NRC ``Kurchatov Institute'' - PNPI), 1 Orlova roscha, Gatchina, 188300 Leningrad region, Russia}
\affiliation{Saint Petersburg State University, 7/9 Universitetskaya nab., St. Petersburg, 199034 Russia}

\author{Alexander V. Oleynichenko}
\email{oleynichenko\_av@pnpi.nrcki.ru \\ alexvoleynichenko@gmail.com}
\affiliation{Petersburg Nuclear Physics Institute named by B.P.\ Konstantinov of National Research Center ``Kurchatov Institute'' (NRC ``Kurchatov Institute'' - PNPI), 1 Orlova roscha, Gatchina, 188300 Leningrad region, Russia}

\author{Andr\'ei V. Zaitsevskii}
\email{zaitsevskii\_av@pnpi.nrcki.ru}
\affiliation{Petersburg Nuclear Physics Institute named by B.P.\ Konstantinov of National Research Center ``Kurchatov Institute'' (NRC ``Kurchatov Institute'' - PNPI), 1 Orlova roscha, Gatchina, 188300 Leningrad region, Russia}
\affiliation{Department of Chemistry, M.V. Lomonosov Moscow State University, Leninskie gory 1/3, Moscow, 119991~Russia}

\begin{abstract}
We estimate the effect of the tensor parity nonconserving (PNC) interaction in the $^{181}$TaO$^+$ molecular cation. It can be used to probe the unknown quadrupole distribution of the neutrons inside the Ta nucleus. To this end, we evaluate the constant which characterizes this interaction using the relativistic Fock space coupled cluster theory for electronic structure modelling. The $^3\Delta_1$ state of the TaO$^+$ cation which can be used to measure the PNC effect is found to be the ground one in agreement with the previous theoretical study.  
\end{abstract}

\maketitle

\section{Introduction}

Experiments with atoms can be used to probe fundamental interactions which violate spatial parity symmetry, $\mathcal{P}$~\cite{Safronova:18}. First experimental evidence of atomic parity non-conservation was obtained in experiments with Bi~\cite{Barko:1978qhi} and Tl~\cite{PhysRevLett.42.343, PhysRevD.24.1134}. Later experimental measurements of the parity nonconserving (PNC) amplitude were reported for Dy~\cite{PhysRevA.56.3453} and Yb~\cite{PhysRevLett.103.071601}. One of the most successful experiments has been carried out on the Cs atom~\cite{Wood:1997}. It allowed one to measure the nuclear spin-independent $\mathcal{P}$-violating effect that arises due to the exchange of Z$^0$-bosons between the nucleus and electrons. The measured effect is in accordance with predictions of the Standard model~\cite{GFreview,Safronova:18}. The experiment has also allowed one to measure for the first time the value of the nuclear anapole moment of Cs.

Significant efforts have been made to investigate the neutron skin effect of the nucleus. For example, such experiments have been recently performed with $^{208}$Pb and $^{48}$Ca isotopes~\cite{Horowitz_Pb_Ca, Prex_2_Pb}. It was shown that the root-mean-square radius of the neutron distribution in $^{208}$Pb is larger than the one of the proton subsystem, which appeared to be the first observation of the neutron skin effect. In these experiments the spherical part of the distribution of neutrons has been studied. However, the quadrupole distribution of neutrons is unknown.
Experimental data about this distribution can give new insight into the nuclear structure. It can be used to test theoretical approaches to describe nuclear and neutron matters.
The nuclear electric quadrupole moment determined by the proton subsystem of the nucleus could be extracted from the hyperfine structure measurement, but such method cannot be used to study the neutron quadrupole distribution. In Ref.~\cite{Flambaum:17} it has been suggested to study the
neutron quadrupole distribution by measuring the tensor contribution to the PNC electron-nucleus interaction. The feature of this interaction is that it is the most sensitive to the quadrupole distribution of the neutron subsystem of the nucleus.

Diatomic molecules are very promising objects to measure PNC effects due to the structure of their energy levels~\cite{Khriplovich:91,KL95,Safronova:18,DeMille:2018,Geddes:18}. In these systems it is possible to find close levels of opposite parity. Recently the experiment with the BaF molecular beam was able to set a constraint on the $\mathcal{P}$-odd nuclear anapole moment of the $^{19}$F nucleus~\cite{DeMille:2018}. Diatomic molecules with a working $^3\Delta_1$ electronic state have been used to set the best constraints on the manifestations of the $\mathcal{T}$,$\mathcal{P}$-odd interactions such as the electron electric dipole moment~\cite{ACME:18}. Molecules in this electronic state have very close levels of opposite parity due to the $\Omega$-doubling effect. Here $\Omega$ is the projection of the total angular momentum of electrons on the internuclear axis of a molecule. Electronic levels with $\Delta\Omega=2$ can be directly mixed by the considered PNC tensor interaction.  In particular, $^3\Delta_{\Omega=1}$ and $^3\Delta_{\Omega=-1}$ states can be mixed. In Ref.~\cite{Fleig:2017} the TaO$^+$ molecular cation has been considered as a candidate system to search for the $\mathcal{T}$,$\mathcal{P}$-violating nuclear magnetic quadrupole moment. Moreover, there was a recent report about experiments on TaO$^+$ formed from the cooled beam of TaO~\cite{ISMS_2021_TaO} molecules. In the present paper we study the effect of mixing of  $^3\Delta_{\Omega=1}$ and $^3\Delta_{\Omega=-1}$ electronic states by the tensor PNC interaction. For this we calculate a molecular constant $W_q$ which characterizes the tensor weak interaction in the $\mathrm{^3 \Delta_1}$ electronic state of $^{181}$TaO$^+$. Using this constant we estimate possible PNC effect. Considered $^{181}$Ta isotope has the nuclear spin $I=7/2$ and the nuclear quadrupole moment $Q=3.17$~b~\cite{STONE20161}. It is stable and the natural abundance of this isotope is 99.988\%, which is favourable from the experimental point of view.

\section{Theory}
The Hamiltonian of the PNC nucleus-electron interaction can be written in the following form~\cite{GFreview}:
\begin{equation}
    h_{PNC} = - \frac{G_F}{2 \sqrt{2}} \gamma_5\{ Z q_{w,p} \rho_p (\mathbf{r})+ N q_{w,n} \rho_n(\mathbf{r}) \}.
\end{equation}
Here $G_F \approx 2.2225 \times 10^{-14}$ a.u. is the Fermi constant, $\gamma_5$ is the Dirac matrix, $Z$ is the number of protons, $N$ is the number of neutrons, $\rho_p (\mathbf{r})$ and $\rho_n(\mathbf{r})$ are densities of protons and neutrons in the nucleus, respectively. $q_{w,p}$ and $q_{w,n}$ are the weak charges of the proton and neutron and have the following values:
\begin{gather}
    q_{w,p} = 1 - 4 \sin^2{\theta_W} \approx 0.08,\\
    q_{w,n} = -1,
\end{gather}
where $\theta_W$ is the Weinberg angle. The application of the multipole expansion for proton and neutron densities while retaining only the first two terms gives~\cite{Flambaum:17}:
\begin{gather}
    \rho_p (\mathbf{r}) \approx \rho_{0p} (r) + \rho_{2p} (r) Y_{20} (\theta, \phi),\\
    \rho_n (\mathbf{r}) \approx \rho_{0n} (r) + \rho_{2n} (r) Y_{20} (\theta, \phi).
\end{gather}
Assuming that $\rho_{0p} = \rho_{0n} = \rho_{0}$ and that $\rho_{2n} (r)$ is proportional to $\rho_0 (r)$, one can arrive at the tensor part of the PNC Hamiltonian~\cite{Flambaum:17}:
\begin{equation}
\label{weakQHam}
    h_Q =- \frac{5 G_F}{2 \sqrt{2} \langle r^2 \rangle} \sum_{q} (-1)^q T_q^{(2)} Q_-q^{TW},
\end{equation}
where $T_q^{(2)} = C_{q}^{(2)} \gamma_5 \rho_0 (r)$ is the electronic part of the operator, $C_{q}^{(2)} = \sqrt{4 \pi/5} Y_{2q}$, $\langle r^2 \rangle = 4 \pi \int \rho_{0} r^4 dr \approx 3 R_{N}^{2} / 5$ is the mean square of nucleus radius, $R_N$ is the nucleus radius. $Q^{TW} = -Q_n + 0.08 Q_p$ is the weak quadrupole moment.

In the present paper we are interested in the following PNC matrix element for the TaO$^+$ cation~\cite{Skripnikov:2019a}:
\begin{equation}
\label{WQme}
     W_q= \langle ^3 \Delta_{+1} | \frac{5 G_F}{2 \sqrt{2} \langle r^2 \rangle} C_{2}^{(2)} \gamma_{5} \rho_{0}(r) | ^3 \Delta_{-1} \rangle.
\end{equation}
This matrix element characterizes the PNC amplitude, which, in principle, can be measured. The value of $W_q$ allows us to estimate possible PNC effect.

\section{Calculation details}

It can be seen, that the action of the electronic part of the operator (\ref{weakQHam}) is concentrated inside the heavy-atom nucleus. At the same time, the required matrix element is determined mainly by the valence part of the electronic wavefunction. Such characteristics are called the ``atom-in-compound'' (AIC) ones~\cite{Titov:14a}. A very efficient method to calculate such characteristics is the two-step approach~\cite{Petrov:02,Titov:06amin,Skripnikov:15b,Skripnikov:16a}. At the first step one performs electronic structure calculation within the generalized relativistic core potential (GRECP) approach~\cite{Titov:99,Mosyagin:10a,Mosyagin:16}. Use of the GRECP operator allows one to decrease the number of basis functions, which are used to expand already ``smoothed'' wavefunctions of valence electrons in the core region of a heavy atom. At the second step of the procedure one uses the non-variational restoration technique~\cite{Petrov:02,Titov:06amin,Skripnikov:15b,Skripnikov:16a} to obtain the correct valence wavefunction in the core region of a heavy atom. This method has been widely used to calculate different AIC characteristics of atoms and molecules~\cite{Skripnikov:2020c,zakharova:2020,Skripnikov:09,Skripnikov:13c,Kudashov:13,Skripnikov:14a,Petrov:13,Skripnikov:17b} and was generalized to solids~\cite{Skripnikov:16a}.

Electronic structure calculations have been performed within the relativistic Fock space coupled cluster method with single and double cluster amplitudes. Sixty electrons of Ta have been excluded from the correlation treatment within the valence version of the GRECP operator constructed in Ref.~\cite{Skripnikov:15c}. Remaining 20 electrons of TaO$^+$ have been included in the correlation calculation. Sector $0h0p$ of the Fock space corresponds to the TaO$^{3+}$ cation. Molecular spinors have been obtained within the relativistic Hartree-Fock method for this cation. Electronic states of interest of TaO$^{+}$ belong to the $0h2p$ sector, i.e. the two particle sector. The model space employed here comprised all distributions of two electrons among up to 16 active molecular spinors. A serious difficulty of the FS-CCSD approach  for such model spaces arises from the presence of intruder states. It leads to small energy denominators in expressions for cluster amplitudes and, hence, numerical instabilities arise during the amplitude equation solution process. To overcome this problem we have used the denominator shifting technique suggested in Refs.~\cite{Zaitsevskii:2017b, Zaitsevskii:2018b}. The following expression was used for the shifted denominators:
\begin{equation}
\label{exprap}
    D_K \longrightarrow D_{K}^{'}(n) = D_K + S_K \Big( \frac{S_K}{D_K+S_K} \Big)^n,
\end{equation}
where $D_K$ is the unshifted energy denominator, $\mathnormal{S_K}$ is the shift parameter for $\mathnormal{D_K}$ and $\mathnormal{n}$ is a non-negative integer attenuation parameter. The shift parameter should be relatively small compared to large negative $\mathnormal{D_K}$ and at the same time be substantial enough so that $\mathnormal{D_K+S_K}$ are all negative and their absolute value is not small. The coupled cluster amplitude equations thus become somewhat approximate, but the corresponding errors can be reduced significantly by varying the $\mathnormal{n}$ parameter and extrapolating the resulting series of effective Hamiltonians to $n=\infty$ (i.e. shifts switched off) using the matrix Pade approximant theory. The employed extrapolation procedure is discussed in details in Ref.~\cite{Zaitsevskii:2018b}.

Calculation of the PNC matrix element~(\ref{WQme}) has been performed within the finite field method. In this method one adds the operator under consideration multiplied by the small parameter $\lambda$ to the electronic Hamiltonian. In the present case this operator mixes two degenerate states leading to the energy shift proportional to the matrix element of the operator between these states and to the parameter $\lambda$. Calculating the numerical derivative one obtains the value of the matrix element. Due to the symmetry properties of the operator and wavefunctions we were not able to use any high symmetry of the system in such approach (the $C_1$ point group have to be considered). To calculate the matrix element we have used the basis set consisting of $15s10p10d5f2g$ uncontracted functions on Ta from Ref.~\cite{Skripnikov:15c}. For oxygen we have employed the Dyall's uncontracted AEDZ~\cite{Dyall:07,Dyall:12,Dyall:2016} basis set. To calculate transition energies of TaO$^{+}$ the basis set on tantalum was increased up to $19s14p12d7f7g4h$ uncontracted functions and up to the uncontracted~\cite{Dyall:07,Dyall:12,Dyall:2016} AAETZ basis set on oxygen. In correlation calculations we have included virtual spinors with energies up to 100 hartree to correctly take into account contribution of the outer core electrons~\cite{Skripnikov:15a,Skripnikov:17a}.

Relativistic Hartree-Fock calculations and integral transformations have been performed within the {\sc dirac}~\cite{DIRAC15,Saue:2020} code.  Coupled cluster calculations were performed within the {\sc exp-t} program system~\cite{Oleynichenko_EXPT,EXPT_website}. To calculate PNC matrix elements the code developed in Refs.~\cite{Skripnikov:2019a,Skripnikov:15b,Skripnikov:11a,Petrov:02} has been used.

\section{Results}

Table~\ref{tabular:tab_examp_2} gives the values of calculated transition energies of TaO$^+$ for several low-lying electronic states.
\begin{table}[h]
\caption{Transition energies of considered states of $\mathrm{TaO^+}$.}
\begin{tabular}{lll}
\hline
\hline
             Term & Occupancy & $\mathnormal{T_e}$, cm$^{-1}$ \\
\hline             
        
             $^3 \Delta_{1}$         &   97\% $\mathrm{ \sigma^{1}_{6s (1/2)_{Ta}} \delta^{1}_{5d (3/2)_{Ta}}}$    &        0       \\        
             $^3 \Delta_{2}$         &
                66\% $\mathrm{ \sigma^{1}_{6s (1/2)_{Ta}} \delta^{1}_{5d (3/2)_{Ta}} }$,  &        1408  \\
                &  31\% $\mathrm{ \sigma^{1}_{6s (1/2)_{Ta}} \delta^{1}_{5d (5/2)_{Ta}}}$  &       \\       
              $^3 \Delta_{3}$         &  97\% $\mathrm{ \sigma^{1}_{6s (1/2)_{Ta}} \delta^{1}_{5d (5/2)_{Ta}}}$    &        3561       \\ 
              $^1 \Sigma_{0}^{+}$     &    59\% $\mathrm{ \sigma^{2}_{6s (1/2)_{Ta}}}$, 36\% $\mathrm{ \delta^{2}_{5d (3/2)_{Ta}}}$      &   3568            \\ 
        
\hline
\hline
\end{tabular}
\label{tabular:tab_examp_2}
\end{table}

It follows from Table~\ref{tabular:tab_examp_2} that the $^3\Delta_1$ state under interest is the ground one in TaO$^+$. Obtained transition energies of low-lying electronic states are in good agreement with the previous theoretical study of this cation~\cite{Fleig:2017}.

\begin{figure}
\caption{Potential energy curves for low-lying electronic states of TaO$^+$.}
\centering
\includegraphics[width = 3.7 in]{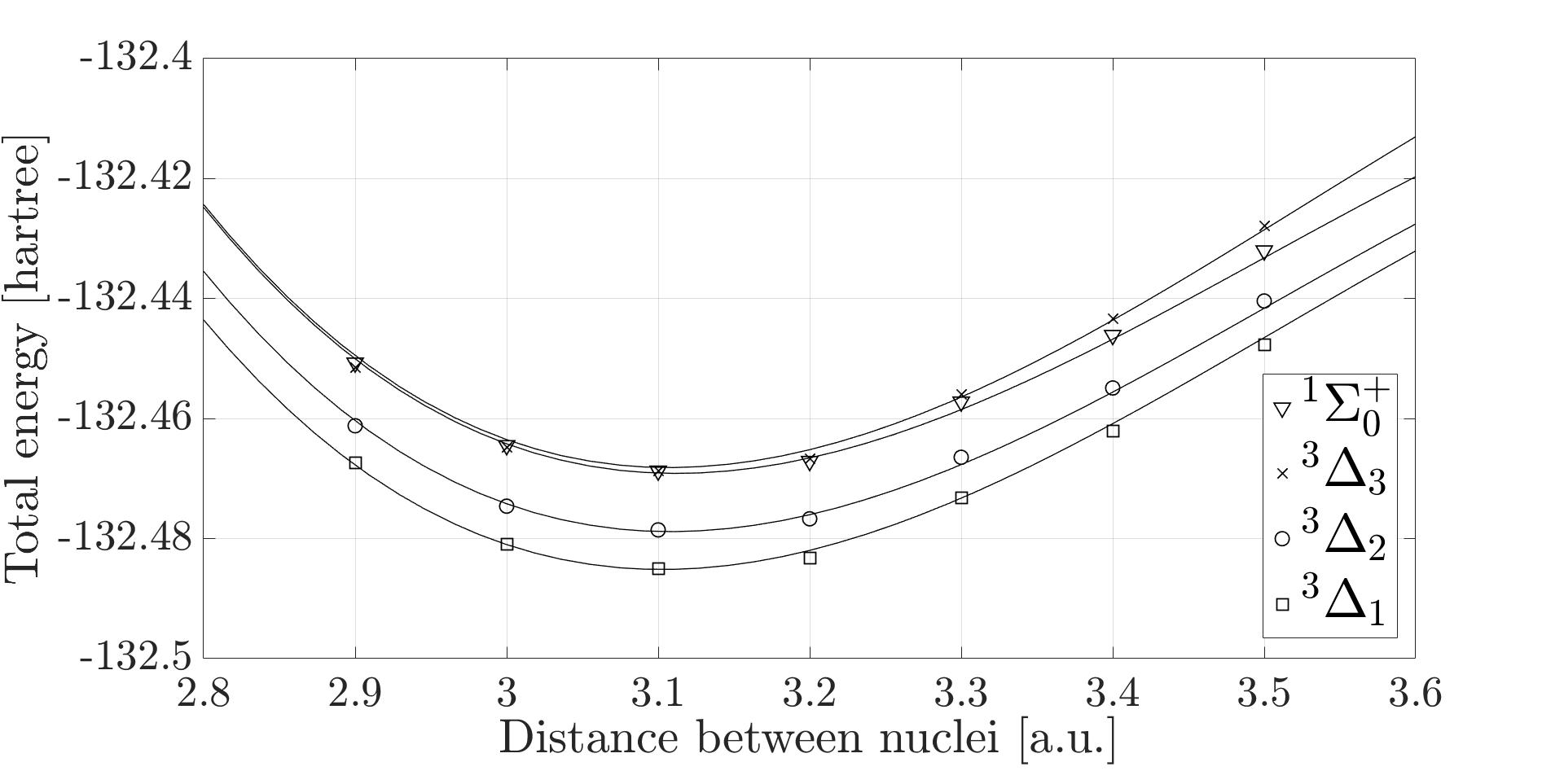}
 \label{FigCurve}
\end{figure}

Potential energy curves of low-lying molecular terms in the vicinity of the equilibrium internuclear distances are given in Figure~\ref{FigCurve}.
Table~\ref{tabular:tab_examp_3} gives theoretical values of the equilibrium distances and harmonic vibrational frequencies.
\begin{table}[h]
\caption{Equilibrium distances $\mathrm{R_e}$ and vibrational frequencies $\mathrm{\omega_e}$ of $\mathrm{TaO^+}$}
\begin{tabular}{llll}
\hline
\hline
Term & Occupancy & $\mathrm{R_e}$, a.u.  & $\mathrm{\omega_e}$, cm$^{-1}$  \\
\hline
$^3 \Delta_{1}$         &  97\% $\mathrm{ \sigma^{1}_{6s (1/2)_{Ta}} \delta^{1}_{5d (3/2)_{Ta}}}$    &        3.11   &   1151    \\
            
            $^3 \Delta_{2}$     &  66\% $\mathrm{ \sigma^{1}_{6s (1/2)_{Ta}} \delta^{1}_{5d (3/2)_{Ta}} }$,  &   3.11     &   1166 \\
            
             &  31\% $\mathrm{ \sigma^{1}_{6s (1/2)_{Ta}} \delta^{1}_{5d (5/2)_{Ta}}}$   &   &   \\
            
            $^3 \Delta_{3}$        &  97\% $\mathrm{ \sigma^{1}_{6s (1/2)_{Ta}} \delta^{1}_{5d (5/2)_{Ta}}}$    & 3.11 &   1194 \\  
            
            $^1 \Sigma^{+}_{0}$        &    59\% $\mathrm{ \sigma^{2}_{6s (1/2)_{Ta}}}, 36\% \mathrm{\delta^{2}_{5d (3/2)_{Ta}}}$      &   3.11     &   1158   \\
\hline
\hline
\end{tabular}
\label{tabular:tab_examp_3}
\end{table}

Calculation of the PNC matrix element (\ref{WQme}) between the $^3\Delta_1$ and $^3\Delta_{-1}$ electronic states is rather challenging. Indeed, in the one-determinant (HF) approximation, the $^3\Delta_1$ state has two unpaired electrons and qualitatively corresponds to $6s_{\sigma, j=-1/2}^1 5d_{\delta, j=+3/2}$ configuration on Ta, while the $^3\Delta_{-1}$ state corresponds to $6s_{\sigma, j=+1/2}^1 5d_{\delta, j=-3/2}$ configuration. Within this approximation, matrix element of any one-particle operator is zero. Thus the value of the considered matrix element~(\ref{WQme}) is completely determined by correlation effects.
We have performed a series of calculations with different numbers of spinors included in the model space to test the convergence of $W_q$. Table~\ref{tabular:tab_examp_4} gives the composition of the Kramers pairs of spinors included in the model space. Each line corresponds to a pair of spinors with opposite signs of $\Omega$.
\begin{table}[h]
 	\caption{Molecular spinors included in the model space in the relativistic FS-CCSD calculations: the values of the projection of the total electronic angular momentum on the molecular axis ($\Omega$),  	one-electron energies ($\mathrm{\varepsilon_{\phi_i}}$) and their leading Mulliken composition.}
\begin{tabular}{llll}
\hline
\hline
 No. & $\mathrm{|\Omega|}$ &  $\mathrm{\varepsilon_{\phi_i}}$, a.u. & Composition  \\
\hline             
10 & 3/2 & -0.854 & 98\% $\mathrm{Ta(d_{\delta})}$ \\
11 & 5/2 & -0.838 & 100\% $\mathrm{Ta(d_{\delta})}$ \\
12 & 1/2 & -0.819 & 74\% $\mathrm{Ta(s)}$, 20\% $\mathrm{Ta(d_{\sigma})}$ \\
13 & 1/2 & -0.768 & 59\% $\mathrm{Ta(d_{\pi})}$, 24\% $\mathrm{O(p_{\pi})}$ \\
14 & 3/2 & -0.757 & 64\% $\mathrm{Ta(d_{\pi})}$, 23\% $\mathrm{O(p_{\pi})}$, 8\% $\mathrm{Ta(p_{\pi})}$ \\
15 & 1/2 & -0.665 & 39\% $\mathrm{Ta(p_{\sigma})}$, 28\% $\mathrm{Ta(d_{\sigma})}$, \\
   &     &        & 13\% $\mathrm{Ta(s)}$, 11\% $\mathrm{O(p_{\sigma})}$ \\
16 & 1/2 & -0.576 & 85\% $\mathrm{Ta(p_{\pi})}$, 5\% $\mathrm{Ta(d_{\pi})}$ \\
17 & 3/2 & -0.563 & 91\% $\mathrm{Ta(p_{\pi})}$, 5\% $\mathrm{Ta(d_{\pi})}$ \\
             
\hline
\hline
\end{tabular}
\label{tabular:tab_examp_4}
\end{table}
Dependencies of the $W_q$ value on the model space size and on the integer parameter $\mathnormal{n}$ of Eq.~(\ref{exprap}) as well as Pade extrapolation procedure result are given in Table~\ref{tabular:tab_examp_5}. 
\begin{table}[h]
 	\caption{Dependence of $W_q$ on considered active Kramers pairs and on the value of denominator shift regularization parameter $\mathnormal{n}$.}
         \begin{tabular}{lll}
         \hline
         \hline
             Active spinors & Shift parameter $\mathnormal{n}$ & $W_q$, $i \times
             10^{-12}$ a.u.  \\
             10..14 & 3 & 2.8   \\
             10..15 & 3 & 4.0   \\
             10..16 & 3 & 3.9  \\
             10..17 & 3 & 3.9  \\
             10..17 & 4 & 4.2  \\
             10..17 & 5 & 4.3  \\
             10..17 & extrapolated & 4.4  \\
             \hline
             \hline
             \end{tabular}
         \label{tabular:tab_examp_5}
 \end{table}
Based on the data from Table~\ref{tabular:tab_examp_5}, it can be validated, that the considered property $W_q$ shows stable behavior when reaching a certain size of the model space. Moreover, $W_q$ is not significantly affected by the value of denominator shifting parameter $\mathnormal{n}$.

Calculated value of the tensor PNC matrix element~(\ref{WQme}) is~$W_q = i \times 4.4 \times 10^{-12}$ a.u.
With this value we can estimate possible PNC effect. Taking the nuclear quadrupole moment of Ta to be $Q_p = 3.17$ b~\cite{STONE20161} and taking into account the fact that for deformed nuclei $Q_n \approx (N/Z) Q_p$~\cite{Flambaum:17}, we estimate the value of $W_q Q^{TW} \approx 5$ mHz. By the order of magnitude this value is close to that found for the HfF$^+$~\cite{Skripnikov:2019a} for the metastable excited $^3\Delta_1$ state. 

As one can see the expected order of magnitude effect is about several mHz. For another molecular experiment on BaF the experimental sensitivity to the nuclear spin-dependent PNC effects was about several Hz ~\cite{DeMille:2018}. At the same time, for the HfF$^+$ cation experiment aimed at measurements of the $\mathcal{T}$,$\mathcal{P}$-violating effects the value for the electron electric dipole moment sensitive frequency channel was less than 1 mHz (statistical)~\cite{Cornell:2017}. Certainly, the experiment to measure the tensor PNC interaction is very challenging, but it will allow one to obtain the neutron quadrupole distribution inside the nucleus for the first time.

\section{Conclusion}

The effect of the tensor electron-nucleus weak interaction was calculated for the $^3\Delta_1$ electronic state of $^{181}$TaO$^+$ molecular cation. This interaction is primarily defined by the neutrons’ quadrupole distribution in the $^{181}$Ta nucleus. The latter is still an unknown property of the nuclear structure. The value of the PNC effect of this interaction was estimated. The $^{181}$TaO$^+$ cation can be rather convenient from the experimental point of view, since the considered working $^3\Delta_1$ term is the ground one. One can note, that in principle other PNC effects such as the nuclear anapole moment and nuclear weak charge can contribute to the considered amplitude through the nonadiabatic effects.
Our previous study of HfF$^+$~\cite{Skripnikov:2019a} showed that such contributions are one to two orders of magnitude less than the effect induced by the
tensor weak interaction. We are going to explore such effects in TaO$^+$ together with possible new physics contributions in future.

\section{Acknowledgments}
\begin{acknowledgments}
    Electronic structure calculations have been carried out using computing resources of the federal collective usage center Complex for Simulation and Data Processing for Mega-science Facilities at National Research Centre ``Kurchatov Institute'', http://ckp.nrcki.ru/, and partly using the computing resources of the quantum chemistry laboratory.

    $~~~$Molecular electronic structure calculations performed at NRC ``Kurchatov Institute'' -- PNPI  have been supported by the Russian Science Foundation Grant No. 19-72-10019. Calculation of matrix elements of PNC interaction performed at SPbU were supported by the foundation for the advancement of theoretical physics and mathematics ``BASIS'' grant according to Project No. 21-1-2-47-1.
\end{acknowledgments}


\end{document}